\documentclass[twocolumn]{aastex631} 

\usepackage[caption=false]{subfig}
\usepackage{pythonhighlight}
\usepackage{float}

\usepackage{amsmath}

\begin{document}

\title{The Atacama Cosmology Telescope: Release of A databaSe of millimeTeR ObservatioNs of Asteroids Using acT ($\mathtt{ASTRONAUT}$)}


\author[0000-0003-3299-3804]{Ricco C. Venterea}
\affiliation{Department of Astronomy, Cornell University, Ithaca, NY 14853, USA}
\affiliation{Cornell Center for Astrophysics and Planetary Science, Cornell University, Ithaca, NY 14853, USA}
\affiliation{Department of Physics and Astronomy, University of California, Riverside, CA 92521, USA}

\author[0000-0003-1842-8104]{John Orlowski-Scherer}
\affiliation{Department of Physics and Astronomy, University of Pennsylvania, 209 South 33rd Street, Philadelphia, PA, USA 19104}

\author[0000-0001-5846-0411]{Nicholas Battaglia}
\affiliation{Department of Astronomy, Cornell University, Ithaca, NY 14853, USA}

\author[0000-0002-4478-7111]{Sigurd Naess}
\affiliation{Institute of Theoretical Astrophysics, University of Oslo, Norway}

\author[0000-0002-9113-7058]{Steve~K.~Choi}
\affiliation{Department of Physics and Astronomy, University of California, Riverside, CA 92521, USA}

\author[0000-0002-7145-1824]{Allen Foster}
\affiliation{Joseph Henry Laboratories of Physics, Jadwin Hall, Princeton University, Princeton, NJ 08544, USA}

\author[0000-0002-4421-0267]{Joseph Golec}
\affiliation{Department of Physics, University of Chicago, Chicago, IL 60637, USA}
\affiliation{Kavli Institute for Cosmological Physics, University of Chicago, Chicago, IL 60637, USA}

\author{Jeffrey J. McMahon}
\affiliation{Department of Physics, University of Chicago, Chicago, IL 60637, USA}
\affiliation{Kavli Institute for Cosmological Physics, University of Chicago, Chicago, IL 60637, USA}
\affiliation{Department of Astronomy and Astrophysics, University of Chicago, Chicago, IL 60637, USA}

\author[0000-0001-6541-9265]{Bruce Partridge}
\affiliation{Department of Physics and Astronomy, Haverford College, Haverford, PA 19041 USA}

\author[0000-0002-8149-1352]{Crist\'{o}bal Sif\'{o}n}
\affiliation{Instituto de F\'{i}sica, Pontificia Universidad Cat\'{o}lica de Valpara\'{i}so, Casilla 4059, Valpara\'{i}so, Chile}

\author[0000-0002-7567-4451]{Edward J. Wollack}
\affiliation{NASA/Goddard Space Flight Center, Greenbelt, Maryland 20771, USA}

\begin{abstract}
We present A databaSe of millimeTeR ObservatioNs of Asteroids Using acT ($\mathtt{ASTRONAUT}$) hosted on Amazon Web Services, Inc. (AWS) in the form of a public Amazon Simple Storage Service (S3) bucket. This bucket is an Amazon cloud storage database containing flux measurements for a group of asteroids at millimeter (mm) wavelengths. These measurements were collected by the Atacama Cosmology Telescope (ACT) from 2017 to 2021 in frequency bands centered near $90$, $150$, and $220$\,GHz. The $\mathtt{ASTRONAUT}$ database contains observation times, normalized flux values, and associated error bars for 170 asteroids above a signal-to-noise ratio of 5 for a single frequency band over the stacked co-added maps. We provide an example in generating light curves with this database. We also present a Jupyter notebook to serve as a reference guide when using the S3 bucket. The container and notebook are publicly available in a GitHub repository.
\end{abstract}


\section{Introduction}\label{sec:intro} 
Studies of asteroids in the solar system have increased our understanding of early planetary formation processes as diverse asteroid spectra have indicated a dynamic solar system history \citep{demeo_2014}. Asteroid measurements at millimeter wavelengths provide a way to study regolith composition since these measurements are sourced from the top of the asteroid subsurface \citep{Ulich_1976, Johnston1982, alma2015}. Different asteroid sub-surface compositions can lead to varying flux measurements as a function of wavelength. Comparing these measurements to models in the infrared (IR) have suggested a sub-millimeter (sub-mm) deficit (e.g., results from \cite{Muller2007, keihm_2013, de_Kleer_2021}). 

Variations in the composition of the asteroid with subsurface depth can signify the presence of temperature gradients. Additionally, longer wavelengths (i.e., in the sub-mm regime) probe deeper in the regolith such that the photon scattering occurring will depend on its composition. Compositional variations will lead to changes in emissivity \citep{Redman1992} allowing multi-wavelength observations to serve as a proxy for regolith composition over depth. Millimeter observations can therefore constrain regolith properties of asteroids. However, there is a lack of millimeter flux data from asteroids compared to other wavelengths. Asteroid observations in the high-frequency radio regime have been around for decades \citep{Conklin1977, redman1995flux, redman1998high, alma2015, de_Kleer_2021, de_Kleer_2024surface}, and most asteroid studies in the millimeter and submillimeter require large observing resources for targeted searches \citep{Muller2007, Chamberlain2007B}. Millimeter observations are also limited by the apparent size of asteroids, as well as their approximately black body spectral radiance profile. Additionally,  asteroids are quite faint at these wavelengths compared to the optical/infrared wavelengths where most asteroid science is done. This is due to both their spectral slope and the diffraction limit imposing a $10^3-10^6$ times worse beam dilution factor.\footnote{Although for interferometric measurements such as \cite{alma2015, de_Kleer_2021, de_Kleer_2024surface}, the angular resolution is higher than the asteroid observations at wavelengths outside of the millimeter so this is not always a recurring limitation.} Millimeter observations are therefore relatively costly, and have until recently been limited to targeted observations of a few asteroids \citep{Chamberlain2007B, Muller2007, alma2015, de_Kleer_2021, li2020}. 

Recent studies have demonstrated the feasibility of collecting asteroid thermal emission flux from cosmology observatories \citep{Chichura2022, Orlowskischerer_2023}. Cosmic microwave background (CMB) survey experiments collect data in the millimeter to sub-millimeter range, which makes telescopes such as the South Pole Telescope (SPT), the Atacama Cosmology Telescope (ACT), and the Simons Observatory (SO) well-suited candidates for studying asteroids in this wavelength regime \citep{abitbol_2025}. Additionally, the CCAT observatory will include a higher frequency range than ACT, while also providing overlapping sky coverage for complementary asteroid spectral information \citep{aravena_2022}. As part of their observations of the CMB, these observatories make repeated observations of large fractions of the sky. This results in incidental observations of large numbers of asteroids but comes at the cost of lower sensitivity compared to past asteroid measurements (e.g., \cite{Muller2007}).

To increase access to these data, we present A databaSe of millimeTeR ObservatioNs of Asteroids Using acT ($\mathtt{ASTRONAUT}$) hosted by Amazon Web Services, Inc. (AWS). This database is in the form of a public Amazon Simple Storage Service (S3) bucket. This bucket contains normalized thermal emission flux measurements, associated error bars, weighting factors based on asteroid geometry, and observation times. To aid in the S3 bucket user implementation, we also include a Jupyter notebook, which is hosted on the $\mathtt{ASTRONAUT}$ GitHub repository. \footnote{\url{https://github.com/ACTCollaboration/ASTRONAUT}} This notebook details methods to pull the flux data and create light curves. We note that the use of the term ``light-curve" in this paper is not standard in the asteroid community. However, this paper is intended for a wider audience, so we keep this usage as it is standard terminology outside of this field and is typically defined as measured flux density as a function of epoch, $S(t)$.

In this paper, we present data products from an analysis of 170 asteroids extracted from observations made by ACT in frequency bands centered near 98, 150, and 228 GHz, denoted f090, f150, and f220, respectively \citep{Orlowskischerer_2023}. ACT can probe the regolith to depths of millimeters to a few centimeters, although these penetration scales into the composite media are directly influenced by wavelength, material properties, composition, and the detailed geometry of the particles in the media \citep{Sihvola_2008, Bohren_2004}. The asteroids presented in this study contain values at the $5$ sigma signal-to-noise (S/N) level or higher in at least one observing band when stacking over all observations. The typical S/N ratio on a single data point taken from a lightcurve can range depending on the asteroid. For brighter asteroids, such as Vesta or Ceres, this ratio can be up to a few while less bright asteroids have $\mathrm{S/N} << 1$.

This paper is structured as follows. In Section~\ref{sec:ACT}, we provide an overview of the ACT data and briefly describe the depth-1 maps used to study asteroids. In Section~\ref{sec:data}, we summarize the asteroid data products from the S3 bucket and describe the bucket itself. We also discuss light curves, the main purpose for this container. In Section~\ref{sec:notes}, we present a Jupyter Notebook which describes how to use the bucket and create an example light curve. Finally, in Section~\ref{sec:conclusion}, we summarize this work.

\section{The Atacama Cosmology Telescope} \label{sec:ACT} 
The Atacama Cosmology Telescope was a $6$\,m off-axis Gregorian telescope located in the Atacama Desert in Chile \citep{Fowler2007, Thornton2016}, which was predominately used for millimeter observations of the cosmic microwave background between 2007 to 2022. ACT made CMB observations in six frequency bands: f$030$, f$040$, f$090$, f$150$, f$220$, and f$280$ GHz \citep{Li2021}. The database presented here contains observations of asteroids from the f$090$, f$150$, and f$220$ GHz channels during 2017-2021. Those three bands are used because maps do not exist for $30$ and $40$ GHz, while $280$ GHz observations were done on small patches of the sky, which makes finding asteroids extremely difficult.
For the expected spectra of asteroids, these three frequency bands are centered around $98$, $150$, and $228$ GHz, with beam full width at half-maximum of $2.0'$, $1.4'$, and $1.0'$, respectively. As ACT uses dichroic detectors, these frequency bands correspond to multiple ACT arrays: pa4, pa5, and pa6. Table \ref{tab:data} indicates the observing frequencies in each array.

\begin{table*}
    \centering    
    \caption{Table of ACT array/frequency combinations, flux density sensitivities, and corresponding beam areas supported by this database. The areas are given at the ACT beam's full width at half maximum \citep{duivenvoorden}. The flux density sensitivity was approximated by finding the variance at the center pixel of each depth-1 map (described in Section \ref{sec:depth1}) for each asteroid and averaging that value across each frequency band. We stress that this flux density sensitivity is only an average and not representative of individual asteroids.}
    \begin{tabular}{cccc}
    \hline\hline\noalign{\smallskip}
        ACT Array & Frequency (GHz) & Sensitivity (mJy) & Beam Area (arcmin$^2$) \\
        \hline
        pa4 & $150$, $220$ & $20$, $73$ & $2.45$, $1.27$ \\
        pa5 & $90$, $150$ & $17$, $20$ & $5.19$, $2.42$ \\
        pa6 & $90$, $150$ & $17$, $20$ & $5.19$, $2.48$ \\
        \hline    
    \end{tabular}
    \label{tab:data}
\end{table*}

\subsection{Depth-1 Maps}
\label{sec:depth1}
ACT maps the sky by scanning at constant elevation while the sky drifts past, resulting in a relatively shallowly exposed image called a ``depth-1" map in ACT Data Release 6, which forms the basis of this analysis \citep{naess_2025}. One can think of depth-1 maps as single-exposure images with an exposure time of around 5 minutes. However, they are not technically single-exposure, as the instrument is continuously read out, and the telescope makes multiple azimuth swipes across a given spot in the sky while it drifts past. The depth-1 maps are made using a modified version of the maximum-likelihood procedure used for the standard ACT maps with fewer conjugate gradient steps \citep{Naess2020}. This saves computational resources at the cost of biasing the maps on large angular scales irrelevant to the point source nature of asteroids. Depth-1 maps are made for each array and associated frequencies of observation.

\subsection{Data Processing}
\label{sec:data_process}
We measure the asteroid fluxes from the depth-1 maps following \cite{Orlowskischerer_2023}. The depth-1 maps are matched filtered for point sources, which cuts out atmospheric noise scales as well as scales smaller than the beam size. This matched-filter maximizes the S/N for point sources on the maps, which includes asteroids. \cite{Orlowskischerer_2023} then create small stamps of depth-1 maps using the JPL Horizons Application Programming Interface \footnote{In general, an API establishes a remote connection between two computers and/or computer programs.} (API).\footnote{\url{https://ssd.jpl.nasa.gov/horizons/}} This API was implemented to determine the location of asteroids on the depth-1 maps using the ACT observation times. Using the Horizons ephemerides, it was possible to interpolate the orbit of asteroids provided observing times. Once these asteroids are centered, measuring the flux corresponds to evaluating the matched-filter flux map value at the location of the asteroid on a single depth-1 map. The flux error is calculated from a flux uncertainty map. \cite{Orlowskischerer_2023} calibrate the data by generating flux maps of Uranus and comparing them to dedicated scans of that planet for ACT \citep{Hajian2011, Hasselfield2013}. Systematic effects, such as the beam size, are also added as an additional uncertainty in the final flux values as described in \cite{Orlowskischerer_2023}. These flux values can be used to create light curves, as described in Section \ref{sec:lcurves}. We note that most asteroids observed by \cite{Orlowskischerer_2023} are noise dominated, meaning that the asteroid signal is less than the noise in a typical depth-1 map. Thus noise fluctuations can produce negative flux in individual data points. This database therefore contains the flux values returned from the center of depth-1 stamps after a normalization procedure, described below. This also includes the flux error, as well as the weighting values used to derive the normalized fluxes. The time corresponds to the observation time of the asteroid, up to the array crossing time (approximately $4$ minutes), which is returned in Unix time units. Unix time is a unit of measurement corresponding to the number of non-leap seconds 
that have passed since 00:00:00 UTC on January 1, 1970 \citep{unix_2013}.  We refer readers to \citet{Orlowskischerer_2023} for more details.


\begin{figure*}
    \centerline{
    \includegraphics[clip,trim=0.0cm 0.0cm 0.0cm 0.0cm,width=\textwidth]{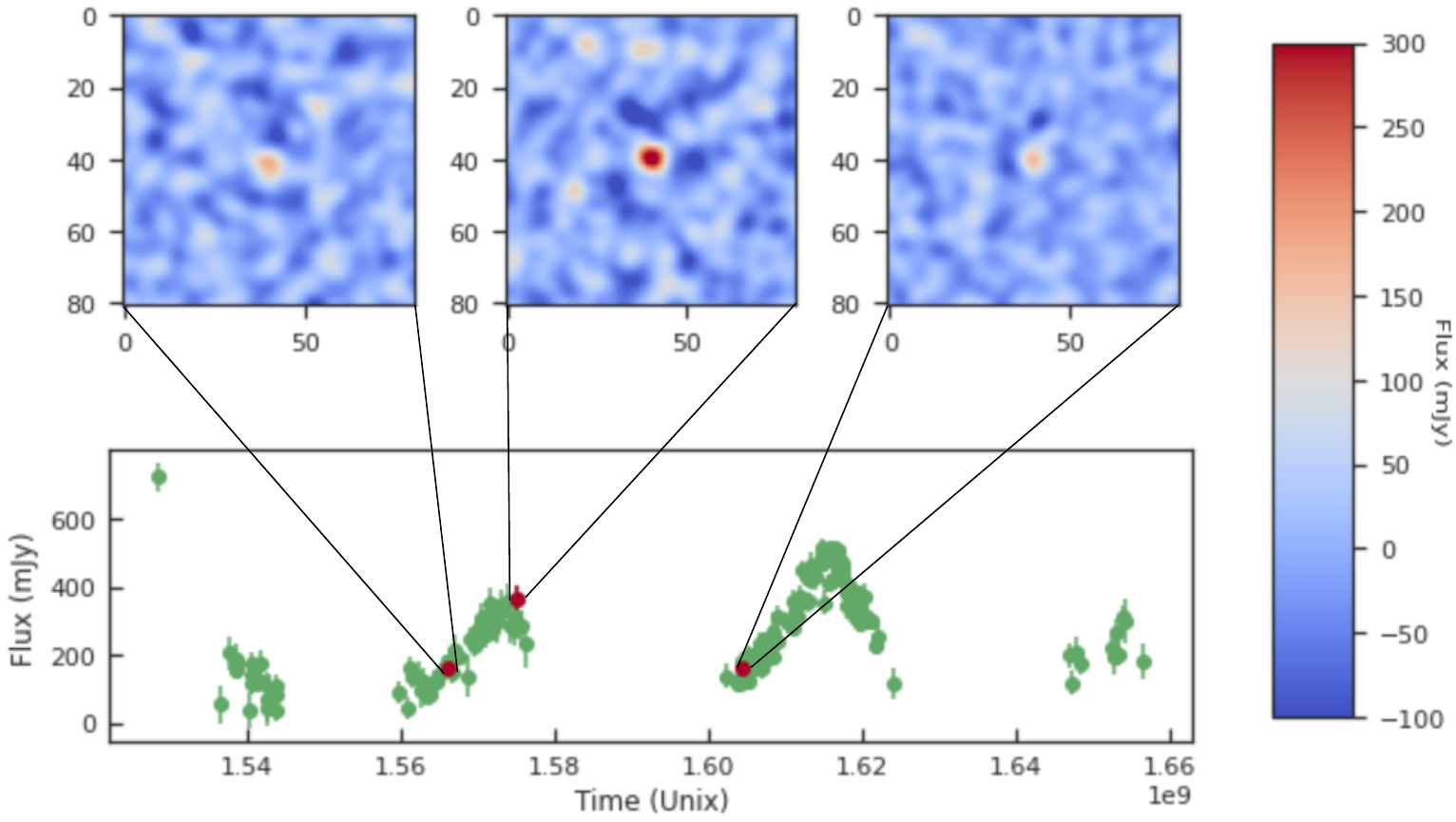}
    }
\caption{Example plot of the depth-1 maps discussed in Section \ref{sec:depth1} which are used to generate light curves discussed in Section \ref{sec:lcurves}. Light curve data shown here are for (4) Vesta on array pa5 at $150$ GHz. The color bar corresponds to flux intensity. Note that this data release does not contain data maps such as those shown here.}
\label{fig:flowchart}
\end{figure*}


\section{Data Products} \label{sec:data}
For each ACT array and frequency band, we provide normalized flux data, flux-errors, weighting factors, and time of observation in the public S3 database. The flux is normalized by the expected flux under the Rayleigh-Jeans limit of the Standard Thermal Model \citep{Lebofsky1986}. We normalize this flux due to the varying asteroid and Earth\//Sun distances between depth-1 maps, as well as the changing observing angle. We can better compare or combine measurements at different geometries through this normalization process. The Rayleigh-Jeans limit can be applied due to the relatively simple spectral energy distribution of asteroids in the far-infrared range \citep{mok_2003}.\footnote{Although this limit gives errors of around $10 \%$ for asteroids observed at these temperatures and upper end of frequencies, this is lower than the uncertainties presented here.} We catalog the flux as follows:
\begin{align} \label{eqn:F_norm}
    F_0 &= F_i (\frac{d_{\mathrm{earth},i}}{1 \mathrm{AU}})^{2}(\frac{d_{\mathrm{sun},i}}{1 \mathrm{AU}})^{1/2}10^{0.004\alpha_i} \\
        &= F_i / W_i
\end{align}
where the weighting scheme $W_i$ is based on the orbital position of the asteroid, which includes the Earth $d_{\mathrm{earth},i}$ and Sun $d_{\mathrm{sun},i}$ centered distances, as well as the Earth-asteroid-Sun angle $\alpha_i$ in degrees. $F_0$ represents the normalized flux for $d_{\mathrm{earth}} = d_{\mathrm{sun}} = 1$ AU. We apply this normalization procedure for each observation stamp $i$. The associated error bars are also normalized in this fashion. The database therefore contains $F_0$, $W_i$, flux error, and observation time. We refer readers to Table \ref{tab:obs} in the Appendix for an overview of the format of the data.

In addition to using the S3 bucket, these data can also be accessed on the Legacy Archive for Microwave Background Data Analysis (LAMBDA) webpage. \footnote{\url{https://lambda.gsfc.nasa.gov/}} The $\mathtt{ASTRONAUT}$ data on LAMBDA are in the form of $\mathtt{tar}$ files. The $\mathtt{tar}$ files provide an alternative to querying this public bucket at the expense of less user-specific file outputs.

While a fuller comparison to existing data was provided in \cite{Orlowskischerer_2023} (to e.g., \cite{Webster1988, Muller2007}), we include the computed ACT flux density measurements hosted in $\mathtt{ASTRONAUT}$ and compare them to results by \cite{redman1998high}. This is seen in a branch of the GitHub repository \footnote{\url{https://github.com/ACTCollaboration/ASTRONAUT/blob/redman1998/Notebooks/Tutorial/Tutorial.ipynb}}, where we compare flux measurements to (1) Ceres, (4) Vesta, (7) Iris, (6) Hebe, and (18) Melpomene. The results are also shown in Table \ref{tab:redman_vs_act}.

\begin{table*} 
    \centering    
    \caption{\cite{redman1998high} flux density comparisons and results made by calling $\mathtt{ASTRONAUT}$. Note that to make comparisons, we first scale the ACT flux measurements to $1$ astronomical unit and then invert Equation (\ref{eqn:F_norm}) to solve for the weighting factor $W_i$. This is then multiplied by the geometric scaling based on the \cite{redman1998high} observation dates presented in Table 3 of that publication. The relevant ephemerides were generated from the JPL Horizons API\tablenotemark{a} at the \cite{redman1998high} epochs.}
    \begin{tabular}{ccccc}
    \hline\hline\noalign{\smallskip}
        Asteroid & arr\_freq & ACT Flux (mJy) & Date (YY/MM) & \cite{redman1998high} Flux (mJy) \\
        \hline
        (1) Ceres & pa5\_f150 & $368.02 \pm 2.27$ & $93/07$ & $463 \pm 58$ \\
        & pa4\_f150 & $379.04 \pm 3.97$ & $93/07$ & $463 \pm 58$ \\
        & pa5\_f150 & $464.42 \pm 2.87$ & $95/05$ & $513 \pm 33$ \\
        & pa5\_f150 & $464.42 \pm 2.87$ & $95/05$ & $513 \pm 33$ \\
        & pa4\_f220 & $797.44 \pm 7.41$ & $93/07$ & $1130 \pm 75$ \\
        (4) Vesta & pa6\_f150 & $368.17 \pm 6.69$ & $93/07$ & $487 \pm 49$ \\
        & pa5\_f150 & $412.96 \pm 2.76$ & $93/07$ & $487 \pm 49$ \\
        & pa4\_f150 & $415.99 \pm 4.78$ & $93/07$ & $487 \pm 49$ \\
        (7) Iris & pa6\_f150 & $203.21 \pm 4.77$ & $93/07$ & $171 \pm 19$ \\
        & pa5\_f150 & $175.72 \pm 3.90$ & $93/07$ & $171 \pm 19$ \\
        & pa4\_f150 & $209.82 \pm 6.01$ & $93/07$ & $171 \pm 19$ \\
        (6) Hebe & pa6\_f150 & $154.83 \pm 12.76$ & $93/07$ & $144 \pm 43$ \\
        & pa5\_f150 & $165.71 \pm 11.48$ & $93/07$ & $144 \pm 43$ \\
        & pa4\_f150 & $174.27 \pm 17.34$ & $93/07$ & $144 \pm 43$ \\
        (18) Melpomene & pa5\_f150 & $85.51 \pm 10.07$ & $93/07$ & $99 \pm 43$ \\
        & pa4\_f150 & $115.38 \pm 18.15$ & $93/07$ & $99 \pm 43$ \\
        & pa4\_f220 & $186.23 \pm 31.97$ & $93/07$ & $279 \pm 50$ \\
        \hline    
    \end{tabular}
    \tablenotetext{a}{\url{https://ssd.jpl.nasa.gov/horizons/}}
    \label{tab:redman_vs_act}
\end{table*}

\subsection{Amazon Simple Storage Service}
\label{sec:API}
A databaSe of millimeTeR ObservatioNs of Asteroids Using acT ($\mathtt{ASTRONAUT}$) is maintained on a public S3 bucket hosted by Amazon's cloud storage service and is located on Amazon's registry of open data on AWS.\footnote{\url{https://registry.opendata.aws/hst/}} The S3 bucket is implemented in a python script. Data can be accessed via python scripts and/or Jupyter Notebooks. 

The bucket returns flux, flux error, weights, and observation times in the form of a $\mathtt{FITS}$ file named $\mathtt{name\_lc\_arr\_freq.fits}$ for each specific query, with $\mathtt{name}$, $\mathtt{arr}$, and $\mathtt{freq}$ being user-defined.

Since the ACT arrays are dichroic, not all frequencies are supported by each array. Therefore, certain queries to the S3 bucket will result in errors if care is not taken (for example, requesting data on array pa6 for frequency $220$ GHz). A result of an invalid request will return $\mathtt{FileNotFoundError}$ to the user.


\subsection{Light Curves}
\label{sec:lcurves}
\begin{figure}
    \centering
    \includegraphics[scale=0.45]{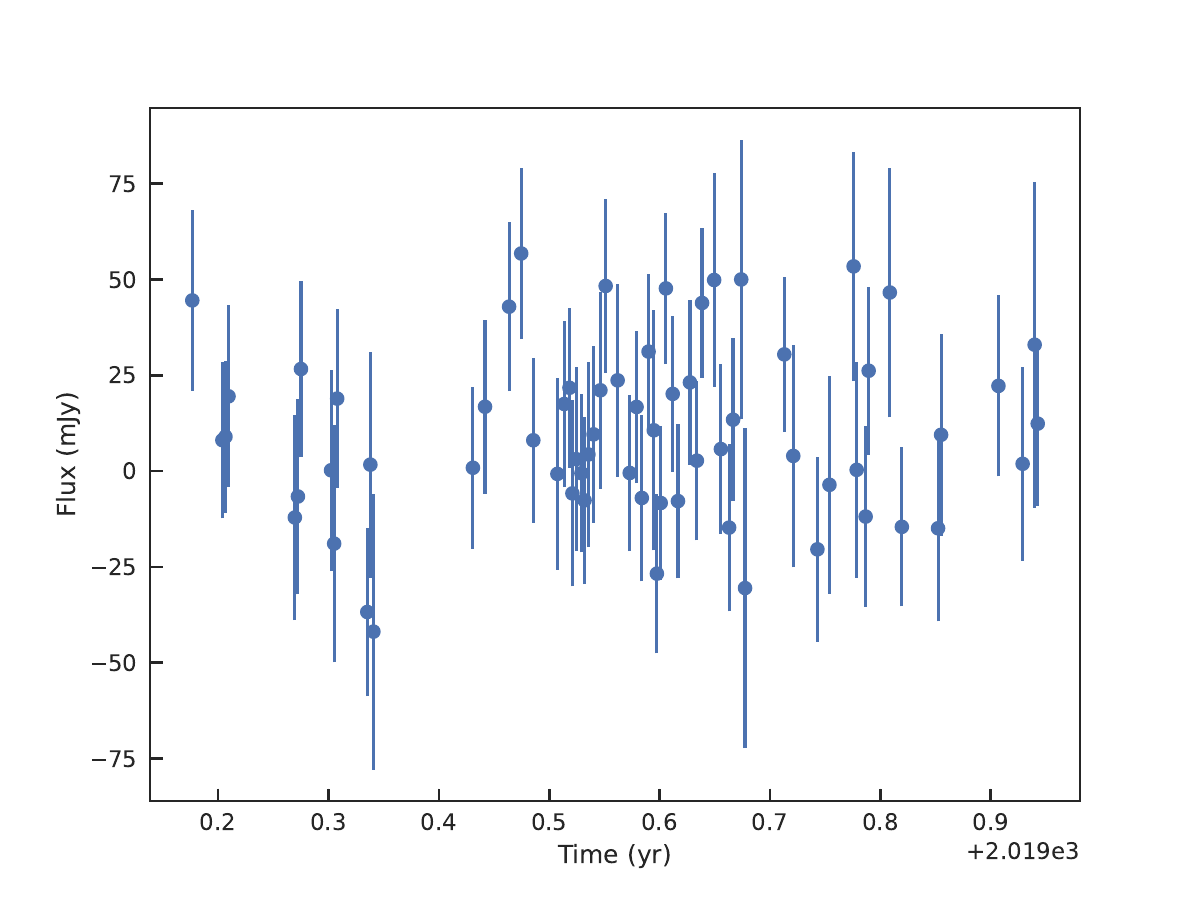}
    \caption{Example light curve at 90 GHz for (705) Erminia using the $\mathtt{ASTRONAUT}$ database. Partial data can be seen in Table \ref{tab:obs} of the Appendix.}
    \label{fig:lcurve}
\end{figure}

By querying this S3 bucket for a specific asteroid, ACT array, and frequency, users can create their own light curve plots. We provide an example plot for (705) Erminia shown in Figure \ref{fig:lcurve}. This plot can be recreated using the code described in the Notebook tutorial. \footnote{\url{https://github.com/ACTCollaboration/ASTRONAUT/blob/main/Notebooks/Tutorial/Tutorial.ipynb}}

\subsection{Available Asteroids}
We also create a script for user-specific requests. For example, users may want a specific asteroid in the database, but may not know if it exists. Using our script with appropriate functions, users may query the $\mathtt{asteroid\_list.txt}$ file for specific requests. We also include functions to search for specific array and frequency combinations. We provide instructions on how to use the script in a markdown file.\footnote{\url{https://github.com/ACTCollaboration/ASTRONAUT/blob/main/lookup.md}}

\section{Jupyter Notebook} \label{sec:notes}
In addition to launching this database, we also create a Jupyter Notebook to aid users. This Notebook guides users through requesting data from the public bucket (described in Section \ref{sec:data}), reading in the data file(s), and generating an example light curve. 

This notebook can be accessed on the $\mathtt{ASTRONAUT}$ GitHub site. \footnote{\url{https://github.com/ACTCollaboration/ASTRONAUT/blob/main/Notebooks/Tutorial/Tutorial.ipynb}} The main purpose of the GitHub is to provide more specific information for queries and host the Jupyter Notebook tutorial. We also provide a file for users to download necessary dependencies.

\subsection{Notebook Tutorial}
\label{sec:Tutorial}
As an example, the notebook describes how to generate asteroid light curves. It starts with users installing relevant python packages, which can be found in the $\mathtt{requirements.txt}$ file in the $\mathtt{ASTRONAUT}$ GitHub. We then describe how users make data requests for specific asteroid(s) on different arrays at different frequencies. We currently support data requests to multiple arrays and frequencies for a specific asteroid. This allows users to generate a light curve across more than one frequency. We then display an example light curve in the Notebook, shown in Figure \ref{fig:lcurve}. 


\section{Conclusion} \label{sec:conclusion}
In this paper, we describe a database containing time, normalized flux, flux uncertainty, and weighting factors for 170 asteroid observations in the millimeter wavelengths. We present this database in the form of a public Amazon Simple Storage Services bucket. These data were collected by the Atacama Cosmology Telescope from 2017-2021. We describe general features of the container, such as data products and basic implementation. We also present a tutorial for users, which can be accessed via Jupyter Notebooks on a GitHub repository. These data are freely available and we encourage the use of these data.

Future applications of these data include creating light curves across observatories, such as ACT, SPT, and the Atacama Large Millimeter/submillimeter Array, as well as calculating phase curves to better understand the variation of millimeter flux as a function of asteroid rotation. 

\section{Acknowledgments} \label{sec:acknowledge}
This work was supported by the U.S. National Science Foundation through awards AST-1440226, AST0965625 and AST-0408698 for the ACT project, as well as awards PHY-1214379 and PHY-0855887. The development of multichroic detectors and lenses was supported by NASA grants NNX13AE56G and NNX14AB58G. ACT operates in the Parque Astron\'{o}mico Atacama in northern Chile under the auspices of the Comisi\'{o}n Nacional de Investigaci\'{o}n Cient\'{i}fica y Tecnol\'{o}gica de Chile (CONICYT), now La Agencia Nacional de Investigaci\'{o}n y Desarrollo (ANID). Colleagues at AstroNorte and RadioSky provide logistical support and keep operations in Chile running smoothly.Computing for ACT was performed using the Princeton Research Computing resources at Princeton University, the National Energy Research Scientific Computing Center (NERSC), and the Niagara and Trillium supercomputers at the SciNet HPC Consortium. SciNet is funded by the CFI under the auspices of Compute Canada, the Government of Ontario, the Ontario Research Fund–Research Excellence, and the University of Toronto.

This work was supported by a grant from the Simons Foundation (CCA 918271, PBL). RCV acknowledges funding and support from the Nexus Scholars Program. Nick Battaglia acknowledges the support from NASA grants 21-ADAP21-0114 and 21-ATP21-0129. 


We thank Bez Thomas and Kailai Wang for their technical support.

\bibliography{main}{}

@ARTICLE{Chamberlain2007B,
       author = {{Chamberlain}, Matthew A. and {Lovell}, Amy J. and {Sykes}, Mark V.},
        title = "{Submillimeter lightcurves of Vesta}",
      journal = {\icarus},
         year = 2007,
        month = dec,
       volume = {192},
       number = {2},
        pages = {448-459},
          doi = {10.1016/j.icarus.2007.08.003},
       adsurl = {https://ui.adsabs.harvard.edu/abs/2007Icar..192..448C}
}

@ARTICLE{Chichura2022,
       author = {{Chichura}, P.~M. and {Foster}, A. and {Patel}, C. and {Ossa-Jaen}, N. and {Ade}, P.~A.~R. and {Ahmed}, Z. and {Anderson}, A.~J. and {Archipley}, M. and {Austermann}, J.~E. and {Avva}, J.~S. and {Balkenhol}, L. and {Barry}, P.~S. and {Thakur}, R. Basu and {Beall}, J.~A. and {Benabed}, K. and {Bender}, A.~N. and {Benson}, B.~A. and {Bianchini}, F. and {Bleem}, L.~E. and {Bouchet}, F.~R. and {Bryant}, L. and {Byrum}, K. and {Carlstrom}, J.~E. and {Carter}, F.~W. and {Cecil}, T.~W. and {Chang}, C.~L. and {Chaubal}, P. and {Chen}, G. and {Chiang}, H.~C. and {Cho}, H. -M. and {Chou}, T. -L. and {Citron}, R. and {Cliche}, J. -F. and {Crawford}, T.~M. and {Crites}, A.~T. and {Cukierman}, A. and {Daley}, C.~M. and {Denison}, E.~V. and {Dibert}, K. and {Ding}, J. and {Dobbs}, M.~A. and {Dutcher}, D. and {Everett}, W. and {Feng}, C. and {Ferguson}, K.~R. and {Fu}, J. and {Galli}, S. and {Gallicchio}, J. and {Gambrel}, A.~E. and {Gardner}, R.~W. and {George}, E.~M. and {Goeckner-Wald}, N. and {Gualtieri}, R. and {Guns}, S. and {Gupta}, N. and {Guyser}, R. and {de Haan}, T. and {Halverson}, N.~W. and {Harke-Hosemann}, A.~H. and {Harrington}, N.~L. and {Henning}, J.~W. and {Hilton}, G.~C. and {Hivon}, E. and {Holder}, G.~P. and {Holzapfel}, W.~L. and {Hood}, J.~C. and {Howe}, D. and {Hrubes}, J.~D. and {Huang}, N. and {Hubmayr}, J. and {Irwin}, K.~D. and {Jeong}, O.~B. and {Jonas}, M. and {Jones}, A. and {Khaire}, T.~S. and {Knox}, L. and {Kofman}, A.~M. and {Korman}, M. and {Kubik}, D.~L. and {Kuhlmann}, S. and {Kuo}, C. -L. and {Lee}, A.~T. and {Leitch}, E.~M. and {Li}, D. and {Lowitz}, A. and {Lu}, C. and {Marrone}, D.~P. and {McMahon}, J.~J. and {Meyer}, S.~S. and {Michalik}, D. and {Millea}, M. and {Mocanu}, L.~M. and {Montgomery}, J. and {Moran}, C. Corbett and {Nadolski}, A. and {Natoli}, T. and {Nguyen}, H. and {Nibarger}, J.~P. and {Noble}, G. and {Novosad}, V. and {Omori}, Y. and {Padin}, S. and {Pan}, Z. and {Paschos}, P. and {Patil}, S. and {Pearson}, J. and {Phadke}, K.~A. and {Posada}, C.~M. and {Prabhu}, K. and {Pryke}, C. and {Quan}, W. and {Rahlin}, A. and {Reichardt}, C.~L. and {Riebel}, D. and {Riedel}, B. and {Rouble}, M. and {Ruhl}, J.~E. and {Saliwanchik}, B.~R. and {Sayre}, J.~T. and {Schaffer}, K.~K. and {Schiappucci}, E. and {Shirokoff}, E. and {Sievers}, C. and {Smecher}, G. and {Sobrin}, J.~A. and {Springmann}, A. and {Stark}, A.~A. and {Stephen}, J. and {Story}, K.~T. and {Suzuki}, A. and {Tandoi}, C. and {Thompson}, K.~L. and {Thorne}, B. and {Tucker}, C. and {Umilta}, C. and {Vale}, L.~R. and {Veach}, T. and {Vieira}, J.~D. and {Wang}, G. and {Whitehorn}, N. and {Wu}, W.~L.~K. and {Yefremenko}, V. and {Yoon}, K.~W. and {Young}, M.~R.},
        title = "{Asteroid Measurements at Millimeter Wavelengths with the South Pole Telescope}",
      journal = {\apj},
         year = 2022,
        month = sep,
       volume = {936},
       number = {2},
          eid = {173},
        pages = {173},
          doi = {10.3847/1538-4357/ac89ec},
archivePrefix = {arXiv},
       eprint = {2202.01406},
 primaryClass = {astro-ph.EP},
       adsurl = {https://ui.adsabs.harvard.edu/abs/2022ApJ...936..173C}
}

@INPROCEEDINGS{Conklin1977,
       author = {{Conklin}, E.~K. and {Ulich}, B.~L. and {Dickel}, J.~R. and {Ther}, D.~T.},
        title = "{Microwave brightness of 1 Ceres and 4 Vesta.}",
    booktitle = {IAU Colloq. 39: Comets, Asteroids, Meteorites: Interrelations, Evolution and Origins},
         year = 1977,
       editor = {{Delsemme}, A.~H.},
        month = jan,
        pages = {257-261},
       adsurl = {https://ui.adsabs.harvard.edu/abs/1977cami.coll..257C}
}

@ARTICLE{Fowler2007,
       author = {{Fowler}, J.~W. and {Niemack}, M.~D. and {Dicker}, S.~R. and {Aboobaker}, A.~M. and {Ade}, P.~A.~R. and {Battistelli}, E.~S. and {Devlin}, M.~J. and {Fisher}, R.~P. and {Halpern}, M. and {Hargrave}, P.~C. and {Hincks}, A.~D. and {Kaul}, M. and {Klein}, J. and {Lau}, J.~M. and {Limon}, M. and {Marriage}, T.~A. and {Mauskopf}, P.~D. and {Page}, L. and {Staggs}, S.~T. and {Swetz}, D.~S. and {Switzer}, E.~R. and {Thornton}, R.~J. and {Tucker}, C.~E.},
        title = "{Optical design of the Atacama Cosmology Telescope and the Millimeter Bolometric Array Camera}",
      journal = {\ao},
         year = 2007,
        month = jun,
       volume = {46},
       number = {17},
        pages = {3444-3454},
          doi = {10.1364/AO.46.003444},
archivePrefix = {arXiv},
       eprint = {astro-ph/0701020},
 primaryClass = {astro-ph},
       adsurl = {https://ui.adsabs.harvard.edu/abs/2007ApOpt..46.3444F}
}

@ARTICLE{Hajian2011,
       author = {{Hajian}, Amir and {Acquaviva}, Viviana and {Ade}, Peter A.~R. and {Aguirre}, Paula and {Amiri}, Mandana and {Appel}, John William and {Barrientos}, L. Felipe and {Battistelli}, Elia S. and {Bond}, John R. and {Brown}, Ben and {Burger}, Bryce and {Chervenak}, Jay and {Das}, Sudeep and {Devlin}, Mark J. and {Dicker}, Simon R. and {Bertrand Doriese}, W. and {Dunkley}, Joanna and {D{\"u}nner}, Rolando and {Essinger-Hileman}, Thomas and {Fisher}, Ryan P. and {Fowler}, Joseph W. and {Halpern}, Mark and {Hasselfield}, Matthew and {Hern{\'a}ndez-Monteagudo}, Carlos and {Hilton}, Gene C. and {Hilton}, Matt and {Hincks}, Adam D. and {Hlozek}, Ren{\'e}e and {Huffenberger}, Kevin M. and {Hughes}, David H. and {Hughes}, John P. and {Infante}, Leopoldo and {Irwin}, Kent D. and {Baptiste Juin}, Jean and {Kaul}, Madhuri and {Klein}, Jeff and {Kosowsky}, Arthur and {Lau}, Judy M. and {Limon}, Michele and {Lin}, Yen-Ting and {Lupton}, Robert H. and {Marriage}, Tobias A. and {Marsden}, Danica and {Mauskopf}, Phil and {Menanteau}, Felipe and {Moodley}, Kavilan and {Moseley}, Harvey and {Netterfield}, Calvin B. and {Niemack}, Michael D. and {Nolta}, Michael R. and {Page}, Lyman A. and {Parker}, Lucas and {Partridge}, Bruce and {Reid}, Beth and {Sehgal}, Neelima and {Sherwin}, Blake D. and {Sievers}, Jon and {Spergel}, David N. and {Staggs}, Suzanne T. and {Swetz}, Daniel S. and {Switzer}, Eric R. and {Thornton}, Robert and {Trac}, Hy and {Tucker}, Carole and {Warne}, Ryan and {Wollack}, Ed and {Zhao}, Yue},
        title = "{The Atacama Cosmology Telescope: Calibration with the Wilkinson Microwave Anisotropy Probe Using Cross-correlations}",
      journal = {\apj},
         year = 2011,
        month = oct,
       volume = {740},
       number = {2},
          eid = {86},
        pages = {86},
          doi = {10.1088/0004-637X/740/2/86},
archivePrefix = {arXiv},
       eprint = {1009.0777},
 primaryClass = {astro-ph.CO},
       adsurl = {https://ui.adsabs.harvard.edu/abs/2011ApJ...740...86H}
}

@ARTICLE{Hasselfield2013,
       author = {{Hasselfield}, Matthew and {Moodley}, Kavilan and {Bond}, J. Richard and {Das}, Sudeep and {Devlin}, Mark J. and {Dunkley}, Joanna and {D{\"u}nner}, Rolando and {Fowler}, Joseph W. and {Gallardo}, Patricio and {Gralla}, Megan B. and {Hajian}, Amir and {Halpern}, Mark and {Hincks}, Adam D. and {Marriage}, Tobias A. and {Marsden}, Danica and {Niemack}, Michael D. and {Nolta}, Michael R. and {Page}, Lyman A. and {Partridge}, Bruce and {Schmitt}, Benjamin L. and {Sehgal}, Neelima and {Sievers}, Jon and {Staggs}, Suzanne T. and {Swetz}, Daniel S. and {Switzer}, Eric R. and {Wollack}, Edward J.},
        title = "{The Atacama Cosmology Telescope: Beam Measurements and the Microwave Brightness Temperatures of Uranus and Saturn}",
      journal = {\apjs},
         year = 2013,
        month = nov,
       volume = {209},
       number = {1},
          eid = {17},
        pages = {17},
          doi = {10.1088/0067-0049/209/1/17},
archivePrefix = {arXiv},
       eprint = {1303.4714},
 primaryClass = {astro-ph.IM},
       adsurl = {https://ui.adsabs.harvard.edu/abs/2013ApJS..209...17H}
}

@ARTICLE{Johnston1982,
       author = {{Johnston}, K.~J. and {Seidelmann}, P.~K. and {Wade}, C.~M.},
        title = "{Observations of 1 Ceres and 2 Pallas at centimeter wavelengths}",
      journal = {\aj},
         year = 1982,
        month = nov,
       volume = {87},
        pages = {1593-1599},
          doi = {10.1086/113249},
       adsurl = {https://ui.adsabs.harvard.edu/abs/1982AJ.....87.1593J}
}

@ARTICLE{Lebofsky1986,
       author = {{Lebofsky}, L.~A. and {Sykes}, M.~V. and {Tedesco}, E.~F. and {Veeder}, G.~J. and {Matson}, D.~L. and {Brown}, R.~H. and {Gradie}, J.~C. and {Feierberg}, M.~A. and {Rudy}, R.~J.},
        title = "{A refined ``standard'' thermal model for asteroids based on observations of 1 Ceres and 2 Pallas}",
      journal = {\icarus},
         year = 1986,
        month = nov,
       volume = {68},
       number = {2},
        pages = {239-251},
          doi = {10.1016/0019-1035(86)90021-7},
       adsurl = {https://ui.adsabs.harvard.edu/abs/1986Icar...68..239L}
}

@ARTICLE{Li2021,
  author={Li, Yaqiong and Austermann, Jason E. and Beall, James A. and Bruno, Sarah Marie and Choi, Steve K. and Cothard, Nicholas F. and Crowley, Kevin T. and Duff, Shannon M. and Ho, Shuay-Pwu Patty and Golec, Joseph E. and Hilton, Gene C. and Hasselfield, Matthew and Hubmayr, Johannes and Koopman, Brian J. and Lungu, Marius and McMahon, Jeff and Niemack, Michael D. and Page, Lyman A. and Salatino, Maria and Simon, Sara M. and Staggs, Suzanne T. and Stevens, Jason R. and Ullom, Joel N. and Vavagiakis, Eve M. and Wang, Yuhan and Wollack, Edward J. and Xu, Zhilei},
  journal={IEEE Transactions on Applied Superconductivity},
  title={In Situ Performance of the Low Frequency Array for Advanced ACTPol},
  year={2021},
  volume={31},
  number={5},
  pages={1-4},

  doi={10.1109/TASC.2021.3063334}}

@ARTICLE{Muller2007,
       author = {{M{\"u}ller}, T.~G. and {Barnes}, P.~J.},
        title = "{3.2 mm lightcurve observations of (4) Vesta and (9) Metis with the Australia Telescope Compact Array}",
      journal = {\aap},
         year = 2007,
        month = may,
       volume = {467},
       number = {2},
        pages = {737-747},
          doi = {10.1051/0004-6361:20066626},
archivePrefix = {arXiv},
       eprint = {astro-ph/0703215},
 primaryClass = {astro-ph},
       adsurl = {https://ui.adsabs.harvard.edu/abs/2007A&A...467..737M}
}

@ARTICLE{Naess2020,
       author = {{Naess}, Sigurd and {Aiola}, Simone and {Austermann}, Jason E. and {Battaglia}, Nick and {Beall}, James A. and {Becker}, Daniel T. and {Bond}, Richard J. and {Calabrese}, Erminia and {Choi}, Steve K. and {Cothard}, Nicholas F. and {Crowley}, Kevin T. and {Darwish}, Omar and {Datta}, Rahul and {Denison}, Edward V. and {Devlin}, Mark and {Duell}, Cody J. and {Duff}, Shannon M. and {Duivenvoorden}, Adriaan J. and {Dunkley}, Jo and {D{\"u}nner}, Rolando and {Fox}, Anna E. and {Gallardo}, Patricio A. and {Halpern}, Mark and {Han}, Dongwon and {Hasselfield}, Matthew and {Hill}, J. Colin and {Hilton}, Gene C. and {Hilton}, Matt and {Hincks}, Adam D. and {Hlo{\v{z}}ek}, Ren{\'e}e and {Ho}, Shuay-Pwu Patty and {Hubmayr}, Johannes and {Huffenberger}, Kevin and {Hughes}, John P. and {Kosowsky}, Arthur B. and {Louis}, Thibaut and {Madhavacheril}, Mathew S. and {McMahon}, Jeff and {Moodley}, Kavilan and {Nati}, Federico and {Nibarger}, John P. and {Niemack}, Michael D. and {Page}, Lyman and {Partridge}, Bruce and {Salatino}, Maria and {Schaan}, Emmanuel and {Schillaci}, Alessandro and {Schmitt}, Benjamin and {Sherwin}, Blake D. and {Sehgal}, Neelima and {Sif{\'o}n}, Crist{\'o}bal and {Spergel}, David and {Staggs}, Suzanne and {Stevens}, Jason and {Storer}, Emilie and {Ullom}, Joel N. and {Vale}, Leila R. and {Van Engelen}, Alexander and {Van Lanen}, Jeff and {Vavagiakis}, Eve M. and {Wollack}, Edward J. and {Xu}, Zhilei},
        title = "{The Atacama Cosmology Telescope: arcminute-resolution maps of 18 000 square degrees of the microwave sky from ACT 2008-2018 data combined with Planck}",
      journal = {\jcap},
         year = 2020,
        month = dec,
       volume = {2020},
       number = {12},
          eid = {046},
        pages = {046},
          doi = {10.1088/1475-7516/2020/12/046},
archivePrefix = {arXiv},
       eprint = {2007.07290},
 primaryClass = {astro-ph.IM},
       adsurl = {https://ui.adsabs.harvard.edu/abs/2020JCAP...12..046N}
}

@ARTICLE{Redman1992,
       author = {{Redman}, Russell O. and {Feldman}, P.~A. and {Matthews}, H.~E. and {Halliday}, I. and {Creutzberg}, F.},
        title = "{Millimeter and Submillimeter Observations of the Asteroid 4 Vesta}",
      journal = {\aj},
         year = 1992,
        month = jul,
       volume = {104},
        pages = {405},
          doi = {10.1086/116248},
       adsurl = {https://ui.adsabs.harvard.edu/abs/1992AJ....104..405R}
}

@article{Thornton2016,
       author = {{Thornton}, R.~J. and {Ade}, P.~A.~R. and {Aiola}, S. and {Angil{\`e}}, F.~E. and {Amiri}, M. and {Beall}, J.~A. and {Becker}, D.~T. and {Cho}, H. -M. and {Choi}, S.~K. and {Corlies}, P. and {Coughlin}, K.~P. and {Datta}, R. and {Devlin}, M.~J. and {Dicker}, S.~R. and {D{\"u}nner}, R. and {Fowler}, J.~W. and {Fox}, A.~E. and {Gallardo}, P.~A. and {Gao}, J. and {Grace}, E. and {Halpern}, M. and {Hasselfield}, M. and {Henderson}, S.~W. and {Hilton}, G.~C. and {Hincks}, A.~D. and {Ho}, S.~P. and {Hubmayr}, J. and {Irwin}, K.~D. and {Klein}, J. and {Koopman}, B. and {Li}, Dale and {Louis}, T. and {Lungu}, M. and {Maurin}, L. and {McMahon}, J. and {Munson}, C.~D. and {Naess}, S. and {Nati}, F. and {Newburgh}, L. and {Nibarger}, J. and {Niemack}, M.~D. and {Niraula}, P. and {Nolta}, M.~R. and {Page}, L.~A. and {Pappas}, C.~G. and {Schillaci}, A. and {Schmitt}, B.~L. and {Sehgal}, N. and {Sievers}, J.~L. and {Simon}, S.~M. and {Staggs}, S.~T. and {Tucker}, C. and {Uehara}, M. and {van Lanen}, J. and {Ward}, J.~T. and {Wollack}, E.~J.},
        title = "{The Atacama Cosmology Telescope: The Polarization-sensitive ACTPol Instrument}",
      journal = {\apjs},
         year = 2016,
        month = dec,
       volume = {227},
       number = {2},
          eid = {21},
        pages = {21},
          doi = {10.3847/1538-4365/227/2/21},
archivePrefix = {arXiv},
       eprint = {1605.06569},
 primaryClass = {astro-ph.IM},
       adsurl = {https://ui.adsabs.harvard.edu/abs/2016ApJS..227...21T}
}

@ARTICLE{Webster1988,
       author = {{Webster}, W.~J. and {Johnston}, K.~J. and {Hobbs}, R.~W. and {Lamphear}, E.~S. and {Wade}, C.~M. and {Lowman}, P.~D. and {Kaplan}, G.~H. and {Seidelmann}, P.~K.},
        title = "{The Microwave Spectrum of the Asteroid Ceres}",
      journal = {\aj},
         year = 1988,
        month = apr,
       volume = {95},
        pages = {1263},
          doi = {10.1086/114722},
       adsurl = {https://ui.adsabs.harvard.edu/abs/1988AJ.....95.1263W}
}

@article{Orlowskischerer_2023,
      title={The Atacama Cosmology Telescope: Millimeter Observations of a Population of Asteroids or: ACTeroids}, 
      author={John Orlowski-Scherer and Ricco Venterea and Nicholas Battaglia and Sigurd Naess and Tanay Bhandarkar and Emily Biermann and Erminia Calabrese and Mark Devlin and Jo Dunkley and Carlos Hervias-Caimapo and Patricio A. Gallardo and Matt Hilton and Adam D. Hincks and Kenda Knowles and Yaqiong Li and Jefferey J. McMahon and Michael D. Niemack and Lyman A. Page and Bruce Partridge and Maria Salatino and Jonathan Sievers and Cristobal Sifon and Suzanne Staggs and Alexander Van Engelen and Cristian Vargas and Eve M. Vavagiakis and Edward J. Wollack},
      year={2024},
    journal={\apj},
    volume={964}, 
    number={2}, 
    doi={10.3847/1538-4357/ad21fe},
    url={https://iopscience.iop.org/article/10.3847/1538-4357/ad21fe/meta}
      
}

@article{demeo_2014,
  title={Solar System evolution from compositional mapping of the asteroid belt},
  author={DeMeo, Francesca E and Carry, Beno{\i}t},
  journal={Nature},
  volume={505},
  number={7485},
  pages={629--634},
  year={2014},
  publisher={Nature Publishing Group UK London}
}

@article{Ulich_1976,
title = {Observations of Ganymede, Callisto, Ceres, Uranus, and Neptune at 3.33 mm wavelength},
journal = {Icarus},
volume = {27},
number = {2},
pages = {183-189},
year = {1976},
issn = {0019-1035},
doi = {https://doi.org/10.1016/0019-1035(76)90001-4},
url = {https://www.sciencedirect.com/science/article/pii/0019103576900014},
author = {B.L. Ulich and E.K. Conklin}
}

@book{unix_2013,
  title={Advanced programming in the UNIX environment},
  author={Stevens, W Richard and Rago, Stephen A},
  year={2013},
  publisher={Addison-Wesley}
}

@article{mok_2003,
  title={Thermal models and far infrared emission of asteroids},
  author={Mok, Lee Hyung and Takao, Nakagawa and Sunao, Hasegawa},
  journal={Journal of the Korean Astronomical Society},
  volume={36},
  number={1},
  pages={21--31},
  year={2003},
  publisher={The Korean Astronomical Society}
}

@article{abitbol_2025,
doi = {10.1088/1475-7516/2025/08/034},
url = {https://doi.org/10.1088/1475-7516/2025/08/034},
year = {2025},
month = {aug},
publisher = {IOP Publishing},
volume = {2025},
number = {08},
pages = {034},
author = {Abitbol, M. and Abril-Cabezas, I. and Adachi, S. and Ade, P. and Adler, A.E. and Agrawal, P. and Aguirre, J. and Ahmed, Z. and Aiola, S. and Alford, T. and Ali, A. and Alonso, D. and Alvarez, M.A. and An, R. and Arnold, K. and Ashton, P. and Atkins, Z. and Austermann, J. and Azzoni, S. and Baccigalupi, C. and Baleato Lizancos, A. and Barron, D. and Barry, P. and Bartlett, J. and Battaglia, N. and Battye, R. and Baxter, E. and Bazarko, A. and Beall, J.A. and Bean, R. and Beck, D. and Beckman, S. and Begin, J. and Beheshti, A. and Beringue, B. and Bhandarkar, T. and Bhimani, S. and Bianchini, F. and Biermann, E. and Biquard, S. and Bixler, B. and Boada, S. and Boettger, D. and Bolliet, B. and Bond, J.R. and Borrill, J. and Borrow, J. and Braithwaite, C. and Brien, T.L.R. and Brown, M.L. and Bruno, S.M. and Bryan, S. and Bustos, R. and Cai, H. and Calabrese, E. and Calafut, V. and Carl, F.M. and Carones, A. and Carron, J. and Challinor, A. and Chanial, P. and Chen, N. and Cheung, K. and Chiang, B. and Chinone, Y. and Chluba, J. and Cho, H.S. and Choi, S.K. and Chu, M. and Clancy, J. and Clark, S.E. and Clarke, P. and Cleary, J. and Clements, D.L. and Connors, J. and Contaldi, C. and Coppi, G. and Corbett, L. and Cothard, N.F. and Coulton, W. and Crowley, K.D. and Crowley, K.T. and Cukierman, A. and D'Ewart, J.M. and Dachlythra, K. and Datta, R. and Day-Weiss, S. and de Haan, T. and Devlin, M. and Di Mascolo, L. and Dicker, S. and Dober, B. and Doux, C. and Dow, P. and Doyle, S. and Duell, C.J. and Duff, S.M. and Duivenvoorden, A.J. and Dunkley, J. and Dutcher, D. and Dünner, R. and Edenton, M. and El Bouhargani, H. and Errard, J. and Fabbian, G. and Fanfani, V. and Farren, G.S. and Fergusson, J. and Ferraro, S. and Flauger, R. and Foster, A. and Freese, K. and Frisch, J.C. and Frolov, A. and Fuller, G. and Galitzki, N. and Gallardo, P.A. and Galvez Ghersi, J.T. and Ganga, K. and Gao, J. and Garrido, X. and Gawiser, E. and Gerbino, M. and Gerras, R. and Giardiello, S. and Gill, A. and Gilles, V. and Giri, U. and Gleave, E. and Gluscevic, V. and Goeckner-Wald, N. and Golec, J.E. and Gordon, S. and Gralla, M. and Gratton, S. and Green, D. and Groh, J.C. and Groppi, C. and Guan, Y. and Gupta, N. and Gudmundsson, J.E. and Hagstotz, S. and Hargrave, P. and Haridas, S. and Harrington, K. and Harrison, I. and Hasegawa, M. and Hasselfield, M. and Haynes, V. and Hazumi, M. and He, A. and Healy, E. and Henderson, S.W. and Hensley, B.S. and Hertig, E. and Hervías-Caimapo, C. and Higuchi, M. and Hill, C.A. and Hill, J.C. and Hilton, G. and Hilton, M. and Hincks, A.D. and Hinshaw, G. and Hložek, R. and Ho, A.Y.Q. and Ho, S. and Ho, S.P. and Hoang, T.D. and Hoh, J. and Hornecker, E. and Hornsby, A.L. and Hotinli, S.C. and Huang, Z. and Huber, Z.B. and Hubmayr, J. and Huffenberger, K. and Hughes, J.P. and Idicherian Lonappan, A. and Ikape, M. and Irwin, K. and Iuliano, J. and Jaffe, A.H. and Jain, B. and Jense, H.T. and Jeong, O. and Johnson, A. and Johnson, B.R. and Johnson, M. and Jones, M. and Jost, B. and Kaneko, D. and Karpel, E.D. and Kasai, Y. and Katayama, N. and Keating, B. and Keller, B. and Keskitalo, R. and Kim, J. and Kisner, T. and Kiuchi, K. and Klein, J. and Knowles, K. and Kofman, A.M. and Koopman, B.J. and Kosowsky, A. and Kou, R. and Krachmalnicoff, N. and Kramer, D. and Krishak, A. and Krolewski, A. and Kusaka, A. and Kusiak, A. and La Plante, P. and La Posta, A. and Laguë, A. and Lashner, J. and Lattanzi, M. and Lee, A. and Lee, E. and Leech, J. and Lessler, C. and Leung, J.S. and Lewis, A. and Li, Y. and Li, Z. and Limon, M. and Lin, L. and Link, M. and Liu, J. and Liu, Y. and Lonergan, J. and Louis, T. and Lucas, T. and Ludlam, M. and Lungu, M. and Lyons, M. and MacCrann, N. and MacInnis, A. and Madhavacheril, M. and Mak, D. and Maldonado, F. and Mallaby-Kay, M. and Manduca, A. and Mangu, A. and Mani, H. and Maniyar, A.S. and Marques, G.A. and Mates, J. and Matsumura, T. and Mauskopf, P. and May, A. and McCallum, N. and McCarrick, H. and McCarthy, F. and McCulloch, M. and McMahon, J. and Meerburg, P.D. and Mehta, Y. and Melin, J. and Meyers, J. and Middleton, A. and Miller, A. and Mirmelstein, M. and Moodley, K. and Moore, J. and Morshed, M. and Morton, T. and Moser, E. and Mroczkowski, T. and Murata, M. and Münchmeyer, M. and Naess, S. and Nakata, H. and Namikawa, T. and Nashimoto, M. and Nati, F. and Natoli, P. and Negrello, M. and Nerval, S.K. and Newburgh, L. and Nguyen, D.V. and Nicola, A. and Niemack, M.D. and Nishino, H. and Nishinomiya, Y. and Orlando, A. and Orlowski-Scherer, J. and Pagano, L. and Page, L.A. and Pandey, S. and Papageorgiou, A. and Paraskevakos, I. and Partridge, B. and Patki, R. and Peel, M. and Perez Sarmiento, K. and Perrotta, F. and Phakathi, P. and Piccirillo, L. and Pierpaoli, E. and Pinsonneault-Marotte, T. and Pisano, G. and Poletti, D. and Puddu, R. and Puglisi, G. and Qu, F.J. and Randall, M.J. and Ranucci, C. and Raum, C. and Reeves, R. and Reichardt, C.L. and Remazeilles, M. and Rephaeli, Y. and Riechers, D. and Robe, J. and Robertson, M.F. and Robertson, N. and Rogers, K. and Rojas, F. and Romero, A. and Rosenberg, E. and Rotti, A. and Rowe, S. and Roy, A. and Sadeh, S. and Sailer, N. and Sakaguri, K. and Sakuma, T. and Sakurai, Y. and Salatino, M. and Sanders, G.H. and Sasaki, D. and Sathyanarayana Rao, M. and Satterthwaite, T.P. and Saunders, L. and Scalcinati, L. and Schaan, E. and Schmitt, B. and Schmittfull, M. and Sehgal, N. and Seibert, J. and Seino, Y. and Seljak, U. and Shaikh, S. and Shaw, E. and Shellard, P. and Sherwin, B. and Shimon, M. and Shroyer, J.E. and Sierra, C. and Sievers, J. and Sifón, C. and Sikhosana, P. and Silva-Feaver, M. and Simon, S.M. and Sinclair, A. and Smith, K. and Sohn, W. and Song, X. and Sonka, R.F. and Spergel, D. and Spisak, J. and Staggs, S.T. and Stein, G. and Stevens, J.R. and Stompor, R. and Storer, E. and Sudiwala, R. and Sugiyama, J. and Surrao, K.M. and Sutariya, S. and Suzuki, A. and Suzuki, J. and Tajima, O. and Takakura, S. and Takeuchi, A. and Tansieri, I. and Taylor, A.C. and Teply, G. and Terasaki, T. and Thomas, A. and Thomas, D.B. and Thornton, R. and Trac, H. and Tsan, T. and Tsang King Sang, E. and Tucker, C. and Ullom, J. and Vacher, L. and Vale, L. and van Engelen, A. and Van Lanen, J. and van Marrewijk, J. and Van Winkle, D.D. and Vargas, C. and Vavagiakis, E.M. and Veenendaal, I. and Vergès, C. and Vissers, M. and Viña, M. and Wagoner, K. and Walker, S. and Walters, L. and Wang, Y. and Westbrook, B. and Williams, J. and Williams, P. and Winch, H. and Wollack, E.J. and Wolz, K. and Wong, J. and Xu, Z. and Yamada, K. and Young, E. and Yu, B. and Yu, C. and Zannoni, M. and Zheng, K. and Zhu, N. and Zonca, A. and Zubeldia, I. and The Simons Observatory collaboration},
title = {The Simons Observatory: science goals and forecasts for the enhanced Large Aperture Telescope},
journal = {Journal of Cosmology and Astroparticle Physics},
}

@article{aravena_2022,
  title={Ccat-prime collaboration: Science goals and forecasts with prime-cam on the fred young submillimeter telescope},
  author={Aravena, Manuel and Austermann, Jason E and Basu, Kaustuv and Battaglia, Nicholas and Beringue, Benjamin and Bertoldi, Frank and Bigiel, Frank and Bond, J Richard and Breysse, Patrick C and Broughton, Colton and others},
  journal={The Astrophysical Journal Supplement Series},
  volume={264},
  number={1},
  pages={7},
  year={2022},
  publisher={IOP Publishing}
}

@article{naess_2025,
  title={The Atacama Cosmology Telescope: DR6 Maps},
  author={Naess, Sigurd and Guan, Yilun and Duivenvoorden, Adriaan J and Hasselfield, Matthew and Wang, Yuhan and Abril-Cabezas, Irene and Addison, Graeme E and Ade, Peter AR and Aiola, Simone and Alford, Tommy and others},
  journal={Journal of Cosmology and Astroparticle Physics},
  volume={2025},
  number={11},
  pages={061},
  year={2025},
  publisher={IOP Publishing}
}

@article{duivenvoorden,
    author = {Duivenvoorden, A.},
    title = {The Atacama Cosmology Telescope: Beam Measurements for DR6},
    journal = {in preparation},
    year = {in preparation}
}

@book{Sihvola_2008,
      author={A. Sihvola},
      title="{Electromagnetic Mixing Formulas and Applications}",
      edition="{Electromagnetic Wave Series}",
      publisher={Institution of Engineering and Technology},
      address={London, United Kingdom},
      year={2008}
}

@book{Bohren_2004,
    author= {Craig F. Bohren and Donald R. Huffman},
    title= "{Light Scattering by Small Particles}",
    year= 2004,
    publisher={Wiley},
    ISBN={0-471-29340-7}
}

@article{redman1998high,
  title={High-quality photometry of asteroids at millimeter and submillimeter wavelengths},
  author={Redman, Russell O and Feldman, PA and Matthews, HE},
  journal={The Astronomical Journal},
  volume={116},
  number={3},
  pages={1478--1490},
  year={1998}
}

@article{alma2015,
  title={The 2014 ALMA long baseline campaign: Observations of asteroid 3 Juno at 60 kilometer resolution},
  author={{ALMA Partnership} and Hunter, TR and Kneissl, R and Moullet, A and Brogan, CL and Fomalont, EB and Vlahakis, C and Asaki, Y and Barkats, D and Dent, WRF and others},
  journal={The Astrophysical Journal Letters},
  volume={808},
  number={1},
  pages={L2},
  year={2015},
  publisher={The American Astronomical Society}
}

@article{keihm_2013,
title = {Reconciling main belt asteroid spectral flux density measurements with a self-consistent thermophysical model},
journal = {Icarus},
volume = {226},
number = {1},
pages = {1086-1102},
year = {2013},
issn = {0019-1035},
doi = {https://doi.org/10.1016/j.icarus.2013.07.005},
url = {https://www.sciencedirect.com/science/article/pii/S0019103513003060},
author = {Stephen Keihm and Lucas Kamp and Samuel Gulkis and Mark Hofstadter and Seungwon Lee and Michael Janssen and Mathieu Choukroun},
}

@article{de_Kleer_2021,
doi = {10.3847/PSJ/ac01ec},
url = {https://doi.org/10.3847/PSJ/ac01ec},
year = {2021},
month = {aug},
publisher = {The American Astronomical Society},
volume = {2},
number = {4},
pages = {149},
author = {de Kleer, Katherine and Cambioni, Saverio and Shepard, Michael},
title = {The Surface of (16) Psyche from Thermal Emission and Polarization Mapping},
journal = {The Planetary Science Journal},
}
\bibliographystyle{aasjournal}

\appendix

\begin{table}[H]
    \centering    
    \caption{The first $10$ observations of (705) Erminia after calling $\mathtt{ASTRONAUT}$.}
    \begin{tabular}{cccc}
    \hline\hline\noalign{\smallskip}
        Time (Unix) & Normalized Flux (mJy) & Flux Error (mJy) & Weight \\
        \hline
        $1.56755315E+09$ & $49.99494371$ & $36.51356554$ & $9.98470761$ \\
        $1.57602406E+09$ & $12.38145675$ & $21.4087249$ & $23.41866536$ \\
        $1.55695744E+09$ & $1.68379052$ & $29.49362796$ & $17.61805998$ \\
        $1.56126720E+09$ & $56.77112365$ & $22.31975483$ & $11.02634292$ \\
        $1.56264640E+09$ & $21.73370641$ & $20.82214211$ & $9.66266889$ \\
        $1.56230093E+09$ & $-0.73528935$ & $25.00771388$ & $9.95554492$ \\
        $1.57076390E+09$ & $53.40798443$ & $29.94858035$ & $14.24851043$ \\
        $1.57559283E+09$ & $1.90102296$ & $25.24328756$ & $22.74014894$ \\
        $1.55479987E+09$ & $-12.10633311$ & $26.76639735$ & $21.34835841$ \\
        $1.55290470E+09$ & $19.51024875$ & $23.71270725$ & $24.3203558$ \\
        \hline    
    \end{tabular}
    \label{tab:obs}
\end{table}

\end{document}